\documentclass{article} % For LaTeX2e
\usepackage{iclr2025_conference,times}

\usepackage{amsmath,amsfonts,bm}

\def\figref#1{figure~\ref{#1}}
\def\secref#1{section~\ref{#1}}
\def\eqref#1{equation~\ref{#1}}
\def\1{\bm{1}}

\def\mA{{\bm{A}}}
\def\mB{{\bm{B}}}

\def\mW{{\bm{W}}}

\DeclareMathAlphabet{\mathsfit}{\encodingdefault}{\sfdefault}{m}{sl}
\SetMathAlphabet{\mathsfit}{bold}{\encodingdefault}{\sfdefault}{bx}{n}

\newcommand{\R}{\mathbb{R}}

\usepackage{hyperref}
\usepackage{url}

\newcounter{notecounter}
\newcommand{\enotesoff}{\long\gdef\enote##1##2{}}
\newcommand{\enoteson}{\long\gdef\enote##1##2{{
\stepcounter{notecounter}
{\large\bf \hspace{1cm}\arabic{notecounter} $<<<$ ##1: ##2 $>>>$\hspace{1cm}}}}}
\enoteson
\enotesoff % uncomment this to turn off the indexed notes

\def\figref#1{Figure~\ref{fig:#1}}
\def\figlabel#1{\label{fig:#1}\label{p:#1}}

\def\tabref#1{Table~\ref{tab:#1}}
\def\tablabel#1{\label{tab:#1}\label{p:#1}}

\def\secref#1{\S\ref{sec:#1}}
\def\seclabel#1{\label{sec:#1}}
\def\eqref#1{Eq.~\ref{eqn:#1}}

\usepackage{tcolorbox}
\usepackage{verbatim}
\usepackage{array}
\usepackage{arydshln}
\usepackage{wrapfig}
\usepackage{booktabs}
\usepackage{enumitem}
\usepackage{svg}
\usepackage{tabularx}
\usepackage{xspace}
\usepackage{subcaption}
\usepackage{graphicx}
\usepackage{amssymb}
\usepackage{amsmath}
\usepackage{relsize}
\usepackage{multirow}
\usepackage{rotating}
\usepackage{mathabx}
\definecolor{cmzhao}{rgb}{0.1, 0.8, 0.1}
\newcommand{\rastrech}[1]{\renewcommand{\arraystretch}{#1}}
\def\mname{OpenMU\xspace}
\def\mul{MU-LLaMA\xspace}
\def\bname{OpenMU-Bench\xspace}

\title{OpenMU: Your Swiss Army Knife for Music Understanding}

\author{\textbf{Mengjie Zhao}\textsuperscript{†}\xspace\xspace
    \textbf{Zhi Zhong}\textsuperscript{†}\xspace\xspace
    \textbf{Zhuoyuan Mao}\textsuperscript{†}\xspace\xspace
    \textbf{Shiqi Yang}\textsuperscript{†}\xspace\xspace \\
    \textbf{Wei-Hsiang Liao}\textsuperscript{‡}\xspace\xspace
    \textbf{Shusuke Takahashi}\textsuperscript{†}\xspace\xspace
    \textbf{Hiromi Wakaki}\textsuperscript{†}\xspace\xspace
    \textbf{Yuki Mitsufuji}\textsuperscript{‡†}\\
    \textsuperscript{†}Sony Group Corporation \hspace{.5cm}
    \textsuperscript{‡}Sony AI\\
}

\iclrfinalcopy % Uncomment for camera-ready version, but NOT for submission.
\begin{document}

\maketitle

\begin{abstract}
   We present \bname, a large-scale benchmark suite for addressing
   the data scarcity issue in training multimodal language models
   to understand music. To construct \bname, we leveraged existing
   datasets and bootstrapped new annotations. \bname also broadens the scope
   of music understanding by including lyrics understanding
   and music tool usage. Using \bname, we trained our music understanding
   model, \mname, with extensive ablations, demonstrating that \mname
   outperforms baseline models such as \mul. Both \mname and \bname are
   open-sourced to facilitate future research in music understanding and to
   enhance creative music production efficiency\footnote{\url{https://github.com/sony/openmu}}.
   \end{abstract}

\section{Introduction}

Multimodal Large Language Models (\textbf{MLLMs}) have successfully extended
large language models (\textbf{LLMs}) by enabling them to perceive, process,
and understand data in modalities beyond text
\citep{tsimpoukelli2021multimodal,liu2023llava,zhu2023minigpt,
mckinzie2024mm1,zhang-etal-2023-video,ltupaper}, such as images,
videos, and audio. 
However, there has been limited effort \citep{gardner2023llark} focused on constructing MLLMs capable of effectively understanding the music modality or addressing Music Information Retrieval (\textbf{MIR}) tasks. 
MIR is a research field focusing on modeling, understanding and interpreting data relevant to music, aiming to improve the efficiency of music production \citep{serra2013roadmap}.
Conventional machine learning algorithms~\citep{Wang2003AnIS,caseymir} sparked the success in music searching. Subsequently, deep learning models expanded the success to music tagging~\citep{won2020cnn}, transcription~\citep{gardner2021mt3, toyama2023hft} and representation learning~\citep{castellon2021calm, li2024mert, won2024foundation}.

We aim to contribute to the MIR field by training an MLLM, 
dubbed \textbf{\mname}, for understanding music clips. 
Building on the versatile capabilities of LLMs and pretrained audio encoders, \mname effectively comprehends and reasons about input music clips, producing relevant answers accordingly. 
We also enable \mname to leverage well-established music tools 
to encourage synergies between \mname and creative practitioners through cooperation. 
\mname is expected to greatly improve music production efficiency. Creative practitioners can instruct \mname to describe a music clip's contents and features, saving minutes of time compared to listening to the full track.

The major obstacle we faced when training and evaluating \mname was the issue of data scarcity in the music modality \citep{serra2013roadmap, seeger2003found, holzapfel2018ethical}.
To address this issue, we construct \textbf{\bname}, a large-scale benchmark for \emph{training and evaluating} MLLMs in music understanding. 
To construct \bname, we bootstrap new datasets using GPT-3.5, and leverage existing datasets when available. 
As a result, \bname comprises approximately one million training examples, covering various aspects of music understanding, 
such as music captioning, reasoning, multi-choice question answering, lyrics understanding and music tool using.
To the best of our knowledge, no large-scale open-sourced benchmark
comparable to \bname currently exists,
and we hope it will advance future research and development of MIR.

In summary, our contributions include: \textbf{(1)}
Proposing \mname for music understanding. \mname is an MLLM dedicated to the music modality, 
outperforming baseline models such as \mul~\citep{mullama} in tasks like music captioning, reasoning, and multiple-choice question answering. 
We carefully evaluate various design choices for \mname and provide extensive ablations on key factors.
\textbf{(2)} Constructing a large-scale benchmark suite, \bname, consisting of approximately one million music understanding data examples. 
We bootstrap new data from GPT-3.5 for rich annotations and also leverage 
existing datasets.
\textbf{(3)} Open-sourcing \mname and \bname. We hope that they will benefit future research and development in music understanding and enhance creative music production by providing rich resources and consistent evaluations.
\vspace{-.3cm}
   
\section{Related work}
\seclabel{sec:relatedwork}
\textbf{Understanding music} goes beyond 
recognizing objective attributes of music 
such as tempo~\citep{bock2015tempo_est, sun2021tempo_est} or 
instrumentation~\citep{gururani2019inst, zhong2023inst}.
It is also subjective and highly context-dependent, like determining music genres~\citep{kereliuk2015gtzan_split}  or moods~\citep{mtgdataset, koutini2019emotion}. 
Researchers succeeded in understanding music by classifying music clips
into predefined tags~\citep{li2024mert, won2024foundation}. 
Recently, \emph{music captioning}~\citep{manco2021muscaps} and \emph{reasoning}~\citep{gardner2023llark} tasks, where natural language descriptions
are employed 
%instead of tags 
to describe music clips, have earned increasing attention. 
Also, the ability of selecting correct answers in \emph{multi-choice question answering}
is included in 
%considered as a part of 
music understanding~\cite{weck2024muchomusic}.
%Despite the growing attention on describe music clips with natural language, 
However,
there has been limited exploration into enabling MLLMs to utilize 
external digital tools (i.e., established music tools) 
for music analysis.
We hypothesize that a music understanding model can further boost 
the workflow of creative practitioners
by deeply integrating the set of widely adopted music tools.
Last but not least, lyrics information processing \citep{watanabe-goto-2020-lyrics}, such as semantic lyrics understanding \citep{lyricsbart} enhances the understanding of a music clip.
%Last but not least, lyrics significantly enhance the understanding of a music clip when present. Lyrics information processing \citep{watanabe-goto-2020-lyrics} tasks, 
%such as semantic lyrics understanding \citep{lyricsbart}, is also critical to music understanding.
Therefore, we integrate it in \bname.
Overall, we \emph{broaden the scope of music understanding} by considering two extra aspects beyond music captioning and reasoning: \emph{Music tool using} and \emph{lyrics understanding}. 

\textbf{Foundation models for music understanding.}
Multimodal LLMs (\textbf{MLLMs}) \citep{tsimpoukelli2021multimodal, liu2023llava, zhu2023minigpt, mckinzie2024mm1, ltupaper}  fuse non-textual information into LLMs \citep{liang-etal-2022-modular} to solve real-world tasks requiring the ability of perceiving data in different modalities.
The scope of MLLMs is recently expanded to include music. \mul~\citep{mullama} and MusiLingo~\citep{deng-etal-2024-musilingo} narrowed down their scope to music captioning and question answering (QA); other critical aspects of music understanding, e.g., key and chord recognition, are not covered.
Perhaps the closest to ours is Llark \citep{gardner2023llark}. However, neither the model itself nor the music understanding datasets from Llark have been released.
None of these models is capable of using music tools, 
an important ability to interact with creators.
In this paper, we propose \textbf{\bname} and \textbf{\mname} to advance the field of music understanding. \bname holistically measures various aspects of music understanding,
%besides music captioning and QA, 
while \mname achieves state-of-art 
%music understanding 
performance on the benchmark.
Both \bname and \mname are released to facilitate the future research and development in this field.
   
\textbf{Music understanding datasets.}
The proliferation of LLMs has spurred the development of benchmarks designed to holistically measure the genuine capabilities of LLMs. 
Benchmarks have been designed for NLP tasks \citep{srivastava2023beyond, hendryckstest2021}, and vision-language tasks~\citep{liu2023mmbench, fu2023mme, ye2023mplug}.
MMMU \citep{yue2023mmmu} included the music modality into evaluation but at a very narrow scope (334 entries of sheet music). 
%due to the data scarcity challenge in music domain.
Researchers are striving to address the data scarcity challenge of music: \citet{lpmusic} introduced LP-MusicCaps, associating LLM-augmented captions with music clips from MusicCaps  \citep{agostinelli2023musiclm}. Similarly, \citet{mullama} developed MusicQA, containing QA and captioning tasks for music clips from MusicCaps, MagnaTagATune \citep{mttdataset}, and MTG-Jamendo \citep{mtgdataset}. Concurrently, \citet{deng-etal-2024-musilingo} proposed MusicInstruct, which targets QA and captioning for clips in MusicCaps. \citet{weck2024muchomusic} create MuChoMusic as a music understanding benchmark containing 1,187 multiple-choice questions for evaluation.
Building on existing datasets, we construct \bname by additionally bootstrapping new datasets using GPT-3.5.  \bname contains about one million examples for training and evaluation across various music understanding tasks.
We also standardize evaluation metrics to ensure consistency\footnote{For example, we found that MU-Llama \citep{mullama} reports BertScore-Recall, while LP-MusicCaps \citep{lpmusic} reports BertScore-F1. We standardize the metrics when reporting performance on \bname, and hope this paves the way for consistent evaluations of music understanding MLLMs.} in reporting results on \bname. \tabref{benchdatastats} provides the statistics for \bname.
   
   \begin{table*}[t]
\centering
\scriptsize
\rastrech{1.2}
\hspace{-.185cm}
\begin{tabular}{r|cc|cc|cc|cc|cc|c|}
\cline{2-12}
\multicolumn{1}{l|}{}                           & \multicolumn{2}{c|}{\textbf{Captioning}} & \multicolumn{2}{c|}{\textbf{Reasoning}} & \multicolumn{2}{c|}{\textbf{Lyrics}} & \multicolumn{2}{c|}{\textbf{Tool-Use}} & \multicolumn{2}{c|}{\textbf{MultipleChoice}} & \multirow{2}{*}{\textbf{Music Clips}} \\ \cline{2-11}
\multicolumn{1}{c|}{}                           & Train               & Test               & Train               & Test              & Train             & Test             & Train              & Test              & Train            & Test            &                                       \\ \hline
\multicolumn{1}{|r|}{\textbf{MusicCaps}}        & 2640                & 2839               & -                   & -                 & -                 & -                & -                  & -                 & -                & -               & 5479                                  \\ \hline
\multicolumn{1}{|r|}{\textbf{MusicInstruct}}    & 28670               & 30593              & -                   & -                 & -                 & -                & -                  & -                 & -                & -               & (5479)                                  \\ \hline
\multicolumn{1}{|r|}{\textbf{LPMusicCaps}}      & 7920                & -                  & -                   & -                 & -                 & -                & -                  & -                 & -                & -               & (2839)                                  \\ \hline
\multicolumn{1}{|r|}{\textbf{LPMusicMTT}}       & 51531               & 13386              & -                   & -                 & -                 & -                & -                  & -                 & -                & -               & 25863                                 \\ \hline
\multicolumn{1}{|r|}{\textbf{Music4all}$^*$}        & 104268              & 5000               & 449711              & 21543             & -                 & -                & -                  & -                 & -                & -               & 109269                                \\ \hline
\multicolumn{1}{|r|}{\textbf{MusicQA-Fin.}} & 31116               & -                  & 38895               & -                 & -                 & -                & -                  & -                 & -                & -               & 12543                                 \\ \hline
\multicolumn{1}{|r|}{\textbf{MusicQA-Test}}     & -                   & 2240               & -                   & 2800              & -                 & -                & -                  & -                 & -                & -               & 500                                   \\ \hline
\multicolumn{1}{|r|}{\textbf{GTZAN}$^*$}            & 639                 & 290                & 2329                & 1116              & -                 & -                & -                  & -                 & -                & -               & 1000                                  \\ \hline
\multicolumn{1}{|r|}{\textbf{MusicNet}$^*$}         & 3791                & 140                & -                   & -                 & -                 & -                & -                  & -                 & -                & -               & 330                                   \\ \hline
\multicolumn{1}{|r|}{\textbf{MTT}$^*$}              & -                   & -                  & 78839               & 16100             & -                 & -                & -                  & -                 & -                & -               & (25863)                                \\ \hline
\multicolumn{1}{|r|}{\textbf{MTG-Jamendo}$^*$}      & 45129               & 5144               & 177771              & 20308             & -                 & -                & -                  & -                 & -                & -               & 50273                                 \\ \hline
\multicolumn{1}{|r|}{\textbf{BART-Fusion}}      & -                   & -                  & -                   & -                 & 55262             & 800              & -                  & -                 & -                & -               & 14985                                 \\ \hline
\multicolumn{1}{|r|}{\textbf{Tool-Using}$^*$}       & -                   & -                  & -                   & -                 & -                 & -                & 1612               & 403               & -                & -               & 0                                     \\ \hline
\multicolumn{1}{|r|}{\textbf{MuChoMusic}}       & -                   & -                  & -                   & -                 & -                 & -                & -                  & -                 & -                & 1187            & 1187                                  \\ \hline
\multicolumn{1}{|r|}{\textbf{Total}}            & 275704              & 54632              & 747545              & 61867             & 55262             & 800              & 1612               & 403               & 0                & 1187            & 221429                                \\ \hline
\end{tabular}
\caption{\bname tasks and dataset distributions. ``MusicQA-Fin.'': MusicQA-Finetuning. $^*$: datasets with our new annotations.
Numbers in brackets are not included when calculating the total number of music clips, as they represent captions annotated for the same set of music clips.}
\tablabel{benchdatastats}
\vspace{-.5cm}
\end{table*}

   \section{Constructing \bname}
   \vspace{-.3cm}
   This section outlines the construction of \bname. We introduce the
   five types of tasks included in \bname and explain the dataset
   construction procedures for each type. In addition to incorporating 
   existing music understanding datasets, we generate new annotations 
   for music clips from datasets that do not contain natural language annotations. 
   Our goal is to integrate as many datasets
   as possible to enable \bname to comprehensively and systematically
   evaluate music understanding models. Furthermore, we specify the
   recommended evaluation metrics to ensure consistent and
   fair benchmarking.
   \tabref{tab:dataexample} shows examples of different \bname task
   types.
   
   \subsection{\bname task types}
   \textbf{Music captioning} tasks a model with generating textual
   descriptions capturing musical contents and key features of a
   music clip. 
   A music understanding model excels at captioning can improve the efficiency of music production by generating music descriptions in a short time, eliminating the needs of listening to the entire music track by creators.
   %A music understanding model excelling at captioning significantly enhances music production efficiency. For example, creators do not have to listen to an entire music track; the model can generate a music clip description by just a single forward pass.
   \textbf{Music reasoning}, as defined by \citet{gardner2023llark},
   tasks the model with answering questions in two aspects. First, it
   examines the interaction between different elements of a music clip,
   such as how a fast tempo is likely to correspond with a high energy
   level. Second, it explores how the real-world can interact with the
   music clip, e.g., how a creator can increase the energy level
   of a music clip by using faster tempos (see \tabref{tab:dataexample}).
   
   \textbf{Tool using}. The MIR community has developed a wide range of
   music technology tools for various tasks, such as tempo estimation,
   key detection, chord recognition, and instrument
   identification\footnote{List of MIR software tools:
   \url{https://www.ismir.net/resources/software-tools/}}. Unlike Llark
   \citep{gardner2023llark}, which aims to address many MIR tasks using
   only the LLM, \mname takes a different view. We aim for \mname to
   integrate and leverage the well-established, rigorously tested MIR
   tools to solve practical, real-world problems. This approach is
   motivated by the limitations of current LLMs, which often struggle
   with tasks like arithmetic \citep{Qin2023ToolLW}.
   By combining the strengths of established MIR tools (e.g., tempo estimator)
   with LLM-based method, \mname aims to provide a more robust solution
   to music understanding.
   
   \textbf{Lyrics understanding}. Lyrics, which carry rich semantic
   content, are often used to convey moods and emotions
   \citep{watanabe-goto-2020-lyrics}. We incorporate a lyrics
   understanding task into \bname. We consider a model that excels at
   understanding lyrics to be capable of producing interpretations
   similar to those of humans, conditioned on the music input. 
   This task is framed as a text generation
   problem.
   %, and we apply the same evaluation metrics used for captioning and reasoning.

   \textbf{Multiple-choice questions}. Text generation models are
   inherently difficult to evaluate due to their open-ended nature
   \citep{celikyilmaz2020evaluation}. This challenge extends to LLM-based
   music understanding models as well. To ensure holistic evaluations,
   \bname also assesses these models using multiple-choice questions
   developed by \citet{weck2024muchomusic}.
   
   \subsection{Individual datasets}
   
   \begin{table*}[t]
\centering
\scriptsize
\rastrech{1.3}
\begin{tabular}{|c|l|}
\hline
\multirow{2}{*}{\textbf{Music Captioning}}      & \textbf{Question}: Write a short   description about the musical contents of this clip.                                                                                                                                                                                                                                                                                                   \\ \cline{2-2}
                                                & \begin{tabular}[c]{@{}l@{}}\textbf{Answer}: The  music clip has a tempo similar to a walking pace and belongs to the  blues genre. It features the \\ characteristic blues  chord progression and soulful  melodies often accompanied by instruments typical \\ of the blues genre are also evident in this clip.\end{tabular}                                                                   \\ \hline
\multirow{2}{*}{\textbf{Music Reasoning}}       & \textbf{Question}: How would a music producer enhance the high energy of this track?                                                                                                                                                                                                                                                                                                              \\ \cline{2-2}
                                                & \begin{tabular}[c]{@{}l@{}}\textbf{Answer}: A music producer could enhance the high energy of this track by adding elements like   driving \\ basslines, powerful drums, and energetic synthesizers. They could   also increase the tempo slightly to create \\ a more intense feel and incorporate dynamic changes in volume and instrumentation to maintain ...\end{tabular} \\ \hline
\multirow{2}{*}{\textbf{Lyrics Understanding}} & \textbf{Question}: Please help me to interpret   this lyrics: \{music\_lyrics\}                                                                                                                                                                                                                                                                                                                    \\ \cline{2-2}
                                                & \textbf{Answer}: Feel like it's about being depressed about your life. You go to work on Monday, see all of the ...                                                                                                                                                                                                                                                              \\ \hline
\multirow{2}{*}{\textbf{Tool Using}}            & \textbf{Question}: What are the chords between 10   sec and 20 sec of this music?                                                                                                                                                                                                                                                                                                                   \\ \cline{2-2}
                                                & \textbf{Answer}: Here are the chords between 10 sec and 20 sec: {[}GetMusicChords(10, 20){]}.                                                                                                                                                                                                                                                                                                   \\\cline{2-2}\hline
\multirow{2}{*}{\textbf{Multiple Choice}}        & \begin{tabular}[c]{@{}l@{}}\textbf{Question}: How would you describe the vocal performance in this piece? \\ Options: (A) Soft yet deeply emotional (B) Lamenting (C) Male vocals (D) Operatic \end{tabular}                                                                   \\\cline{2-2}
                                                & \textbf{Answer}: (A) Soft yet deeply emotional.                                                                                                                                                                                                                                                                                                   \\ \hline
\end{tabular}
\caption{Example data entries in \bname.}
\tablabel{tab:dataexample}
\vspace{-0.5cm}
\end{table*}

   As introduced in \secref{sec:relatedwork}, a few datasets already exist for music understanding. 
   We incorporate these datasets and create new annotations to ensure that \bname has both a large scale and broad coverage. 
   We describe each of the datasets, along with the applied modifications aligning them with the \bname task types.
   We adhere to existing train/test splits of the datasets 
   when available (c.f. \secref{apdx:splits});
   Appendix \secref{apdx:metadata} details the preprocessing and annotating details
   of \bname; we highlight only the key information here.
   
   \textbf{MusicCaps}, created by \citet{agostinelli2023musiclm}, is
   pivotal for the music captioning task. It contains approximately 5.5K
   10-second music clips sourced from AudioSet \citep{gemmeke2017audio},
   with corresponding gold-standard text captions written by professional
   musicians. We incorporate MusicCaps into \bname as part of the
   captioning task.
   \textbf{LPMusicCaps \& LPMusicMTT} \citep{lpmusic} extend MusicCaps
   and the MagnaTagATune
   \citep{MTT} dataset by generating additional textual descriptions.
   The authors prompt GPT-3.5 to ``write'', ``summarize'', ``paraphrase'',
   and ``predict attributes'' new captions to the music clips.
   We integrate\footnote{We do not use the ``attribute prediction''
   annotations, following the recommendation from the LPMusic authors:
   \url{https://huggingface.co/datasets/seungheondoh/LP-MusicCaps-MC}}
   approximately 8K LPMusicCaps and 51K LPMusicMTT training captions into \bname.
   \textbf{MusicInstruct} \citep{deng-etal-2024-musilingo} also extends
   MusicCaps by creating question-answer pairs for the MusicCaps clips
   using GPT-4. This dataset contains approximately 60K question-answer
   pairs, which are categorized into two versions: a short version (\textbf{MI-short})
   focusing on musical content such as tempo and genre, and a long
   version (\textbf{MI-long}) that paraphrases the MusicCaps captions. We integrate
   MusicInstruct into \bname as a captioning task, and report performance
   on both versions separately.
   
   \textbf{MusicQA}, developed by \citet{mullama} by prompting MPT
   \citep{MosaicML2023Introducing}, is employed to train their
   \mul. MusicQA is composed of MusicCaps clips for pretraining,
   MagnaTagATune \citep{MTT} clips for finetuning, and MTG-Jamendo
   \citep{mtgdataset} clips for testing. We incorporate MusicQA-Finetune
   and MusicQA-Test into \bname, while MusicQA-Pretrain, which contains
   the test split of MusicCaps, is excluded to prevent potential
   train-test leakage \citep{deng-etal-2024-musilingo}. Following
   \citet{mullama}, we separate MusicQA into captioning and reasoning
   parts.

   \textbf{Music4all}, developed by \citet{music4all}, consists of
   approximately 100K music clips with rich metadata, including
   attributes like energy, valence, and genre. Based on this metadata, we
   prompt GPT-3.5 to generate annotations for both the captioning and
   reasoning tasks. The prompts used for these annotations are provided
   in the Appendix \secref{apdx:prompts}.
   \textbf{GTZAN}, developed by \citet{GTZAN}, contains approximately 1K
   30-second music clips, each labeled with genre tags 
   and we create extra tempo tags with \texttt{Madmom} \citep{madmom}. 
   Based on these tags, we generate captioning and reasoning annotations with prompting.
   \textbf{MusicNet} \citep{musicnet} contains 1 million dense
   annotations at precise timestamps for 330 classical music
   recordings. The annotations are of high quality, but primarily focus
   on instruments. As a result, we integrate MusicNet into \bname as part
   of the captioning task, retaining only annotations that span three
   seconds or longer.
   \textbf{MagnaTagATune} (MTT) has been included in \bname as part of
   the captioning task, thanks to the annotations by
   \citet{lpmusic}. Given its significance in the MIR community~\citep{won2020cnn}, we also
   create an additional 90K reasoning annotations for training and
   testing.
   \textbf{MTG-Jamendo} \citep{mtgdataset} consists of approximately
   55K full music tracks, each tagged with genre, instrument, and mood.
   We randomly select 30-second music clips\footnote{We provide scripts 
   for extracting music clips identical to ours.} and generate
   annotations for captioning and reasoning tasks by prompting GPT-3.5.
   
   \textbf{Tool using}. To the best of our knowledge, there is no
   existing dataset designed to train models in leveraging MIR tools. To
   address this, we generate training and testing
   datasets for solving four MIR tasks with tools: chord recognition,
   tempo estimation, key detection, and downbeat extraction. We
   demonstrate that \mname quickly learns to utilize these tools to
   answer queries related to MIR information.
   We implemented these tools by wrapping the Python package
   \texttt{Madmom} \citep{madmom}; 
   \secref{apdx:tools} shows implementation details.

   For \textbf{Lyrics understanding}, we integrate BART-fusion
   \citep{lyricsbart}'s annotations, containing internet
   interpretations to the lyrics and music clips of Music4all.  For
   \textbf{Multiple-choice questions}, we integrate MuChoMusic
   \citep{weck2024muchomusic} for evaluation. The task involves answering
   questions about music knowledge and reasoning by selecting the
   correct option from four provided choices.

   \subsection{Evaluation metrics}  \bname leverages common evaluation
   metrics for text generation tasks: captioning, reasoning, and
   lyrics understanding. BLEU-1, BLEU
   \citep{papineni-etal-2002-bleu}\footnote{Following the machine translation literature, our BLEU refers to BLEU-4.}, 
   %(Following the machine translation literature, our BLEU refers to BLEU-4) %comment: this does not save much space
   Meteor
   \citep{banerjee-lavie-2005-meteor}, Rouge-1, and Rouge-L
   \citep{lin-2004-rouge} measure an answer's textual overlap with
   the gold standard, while BertScore \citep{bertscore} measures
   similarity in the semantic representation space of a pretrained BERT model. For all evaluations, we report
   the scores computed using the F-measure.
   We report accuracy for the task of multiple-choice questions.
   
   \section{Model architecture and training details}
   \vspace{-.3cm}
   \subsection{Model architecture}
   \textbf{Encoding music clips.} We use AudioMAE
   \citep{huang2022masked} to encode an input music clip into vector
   representations. Specifically, we use the ``ViT-B AS-2M pretrained +
   finetuned'' version of AudioMAE, which is a Vision Transformer \citep{dosovitskiy2021an} initially pretrained with a masked auto-encoding reconstruction loss \citep{he2022masked}, followed by
   finetuning on tagging tasks \citep{gemmeke2017audio}, both using the AudioSet2M \citep{gemmeke2017audio} dataset.
   The choice of using AudioMAE over other music encoders, such as MERT \citep{li2024mert} or Jukebox-5B
   \citep{dhariwal2020jukebox, castellon2021calm}, is motivated by two primary
   reasons. 
   First, more than half of the audio clips inAudioSet2M\footnote{AudioSet2M Ontology:
   \url{https://research.google.com/audioset/}} consist of music or musical instrument recordings, resulting in approximately 3,137 hours
   of music data (compared to the 910 hours in the MERT-95M-public model \citep{li2024mert}).
   Audio encoders pretrained on AudioSet have shown competitive performance in music tagging tasks~\citep{koutini2021passt, niizumi2022audiomae}.
   Second, the size of the multimodal encoder is not
   a performance bottleneck \citep{mckinzie2024mm1}. Instead, the smaller
   number of parameters in ViT-B (86M) facilitates more efficient training.
   
   \textbf{LLM}. We use the open-sourced
   Llama3-8B-instruct \citep{llama3} as our LLM. Compared to previous
   Llama models \citep{touvron2023llama}, Llama3 has been trained on
   higher-quality datasets and at larger scales, achieving GPT-4-level
   performance \citep{achiam2023gpt} on numerous tasks.
   
   \textbf{Music-language projector} links the
   representation space of the music encoder with the LLM. Studies~\citep{mckinzie2024mm1,liu2023improvedllava} have shown that the
   architecture of the projector itself has little impact on downstream
   task performance, while the number of tokens from the multimodal
   encoder is significantly important. We use a two-layer MLP with GELU
   non-linearity \citep{gelu} and evaluate the effect of varying the
   number of music tokens in \secref{sec:design:tokens}.
   
   Overall, \mname follows the well-tested architecture of MLLMs \citep{mckinzie2024mm1}
   as shown in \figref{modelarchi}. In contrast to previous music MLLMs such as
   \mul, \mname is also capable of interacting with external MIR tools
   such as tempo estimator \citep{madmom}.
   
   \begin{figure}[t]
   \centering
   \hspace{.8cm}\includegraphics[width=.7\textwidth]{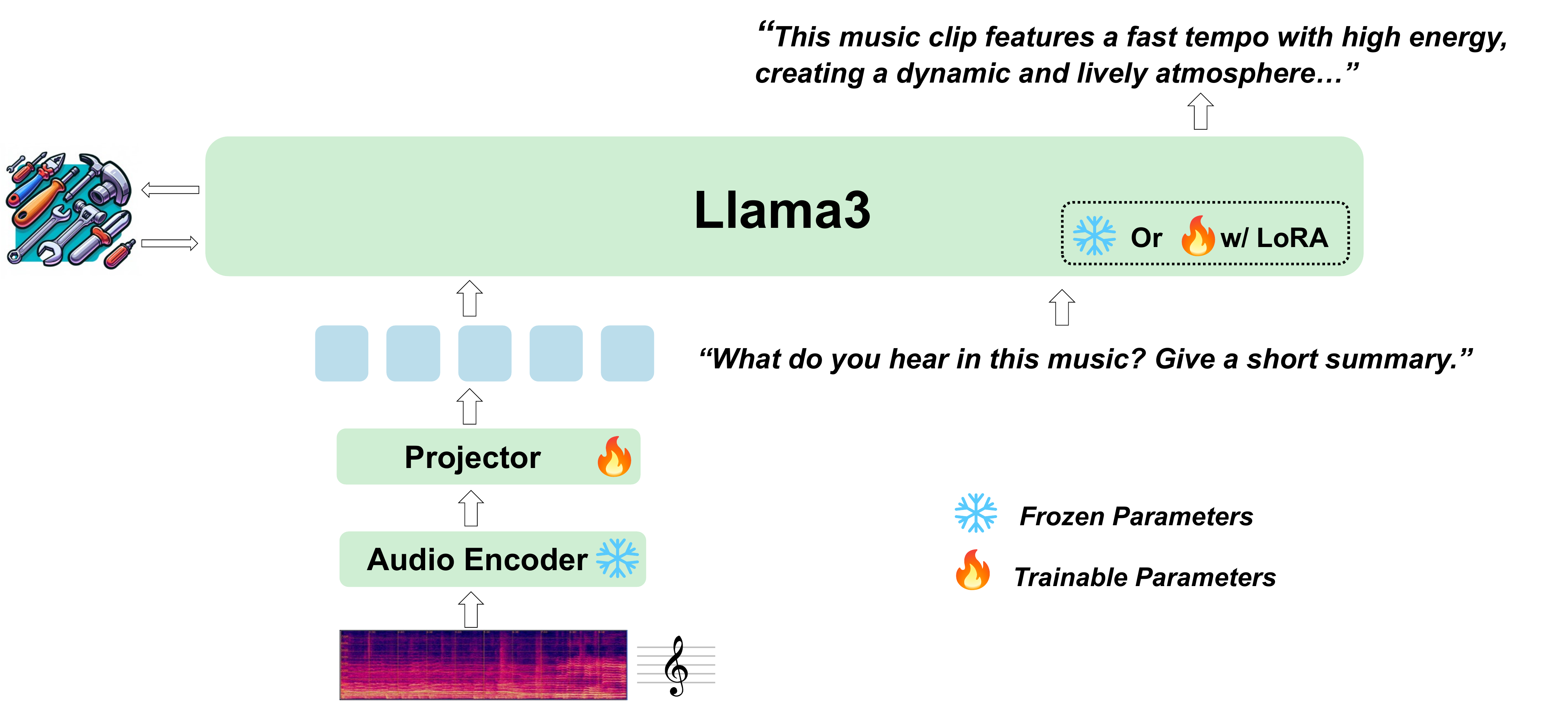}
   \caption{Model architecture of \mname. In Stage (1), we only tune the
   music-language projector. In Stage (2), LoRA adapters are added to the
   LLM and are tuned together with the projector.}
   \figlabel{modelarchi}
   \vspace{-.5cm}
   \end{figure}
   
   \subsection{Training details}
   \seclabel{sec:traindetails}
   
   \textbf{Dataset preprocessing}. When processing the music clips, we
   limit their maximum length to 30 seconds and zero-pad those shorter
   than 30 seconds. 
   All music clips are resampled to 16 kHz and then converted to a 128-bin Mel-spectrogram with a 25-ms hann window and 10-ms hop size.
   %All music clips are resampled to 16 kHz, and their waveforms are converted into mel-spectrograms. The number of mel-frequency bins is set to 128. We use a Hanning window with a frame length of 25 milliseconds and a frame shift of 10 milliseconds. 
   Consequently, each music clip is represented as a
   mel-spectrogram with a shape of (3072, 128). Since AudioMAE is trained to
   encode inputs of up to 10 seconds, we segment each mel-spectrogram
   into three parts, encode them separately, and then concatenate the
   results. As a result, each 30-second music clip is encoded by
   1536 tokens, with each token having a shape of (1, 768).
   
   Throughout our experiments, we used between 8 and 16 A100 40GB GPUs,
   depending on the experimental setup (c.f. \secref{sec:design}). In all
   experiments, we set the maximum context length of the LLM to 2048
   tokens. We utilized DeepSpeed ZeRO-3 \citep{rajbhandari2020zero} and
   FlashAttention2 \citep{dao2022flashattention} to enable fast and
   efficient training. It took approximately three days to train \mname on
   the captioning and reasoning subsets of \bname (around 1 million data
   examples).
   
   \textbf{Training setup} of \bname largely follows the common practice
   of MLLM training \citep{yin2023survey,liu2023llava,mckinzie2024mm1},
   consisting of:
   
   \begin{itemize}[leftmargin=5mm]
   \renewcommand\labelitemi{}
   
   \item \textbf{Stage (1) Captioning}. We train \mname to generate
   captions, conditioned solely on the input music clip. The goal of
   Stage (1) training is to align the representation spaces of AudioMAE and
   Llama3, with the only trainable module in this stage being the
   music-language projector. We use the captioning subset of \bname for
   training in this stage. A key configuration is the number of music
   tokens fed into the LLM, which we discuss in detail in
   \secref{sec:design:tokens}.
   The remaining hyperparameters largely follow
   \citet{liu2023llava} and are provided in the Appendix \secref{apdx:hypers}.
   
   \item \textbf{Stage (2) Instruction Tuning}. After aligning the music
   and text representation spaces, Stage (2) training enables \mname to
   follow various instructions in the music domain, such as inferring
   music genres or reasoning about the content of a music clip. In this
   stage, LoRA adapters \citep{hu2022lora} are incorporated into \mname's
   LLM, followed by fine-tuning on \bname's captioning and reasoning tasks. We focus on
   two critical research questions in this stage. First, we extensively
   evaluate \mname's task performance with respect to its LoRA parameters
   (see \secref{sec:design:lorataskhallu}). 
   Second, we investigate in-depth \mname's use of music information. 
   Given the large-scale pretraining data of \mname's LLM, we hypothesize that \mname might be able to make correct predictions for knowledge-intensive questions even without relying on musical information within a music clip. 
   We test this hypothesis and show that in order to achieve higher performance, \mname indeed relies on information from the music clip, demonstrating \mname's genuine ability to understand music.
   \end{itemize}

   \begin{figure}[t]
   \centering
   \includegraphics[width=0.9\textwidth,height=0.2\textheight]{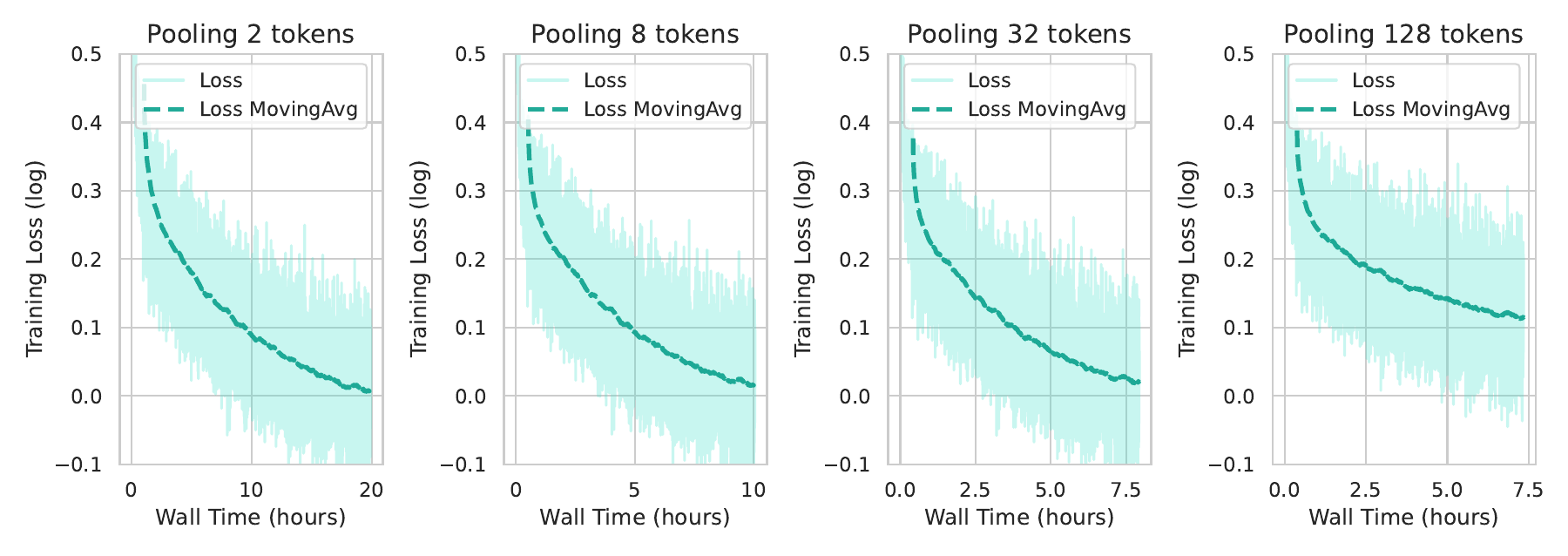}\\
   \hspace{.5cm}
   \includegraphics[width=0.9\textwidth,height=0.2\textheight]{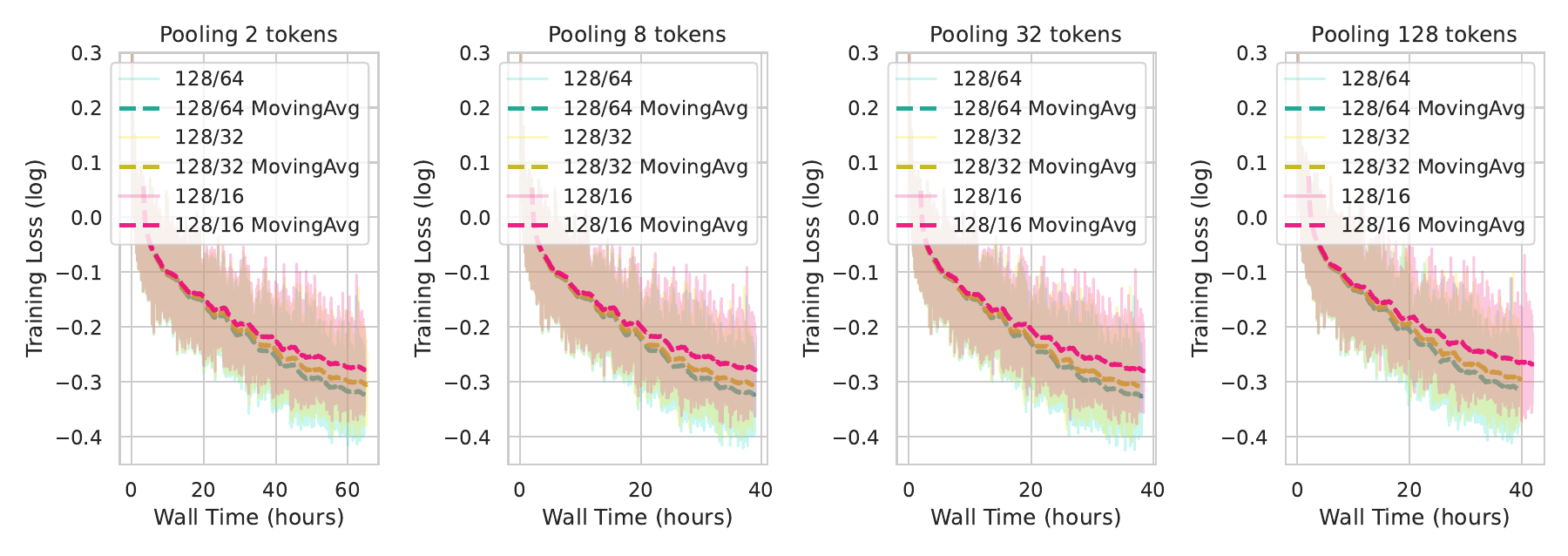}
   \caption{ Training trajectories of Stage (1) (top) and Stage (2)
   (bottom). The x-axis represents the number of hours elapsed, and the
   y-axis shows the training loss on a log scale. We vary the number of
   mean-pooling music tokens from 2 to 128 and experiment with different
   LoRA parameter combinations, $\alpha/r$. ``MovingAvg''
   represents the moving average.}
   \figlabel{img:tokens}
   \end{figure}
   
   \section{Designing \mname and discussions}
   \vspace{-.3cm}
   \seclabel{sec:design}
   In this section, we explore and discuss the critical factors involved
   in training \mname. We aim for these detailed analyses to contribute
   to the research and development of future foundation models for music
   understanding.  

   \subsection{Number of music tokens}
   \seclabel{sec:design:tokens}
   \citet{mckinzie2024mm1} illustrate that the number of image tokens is
   more significant than the architecture of the vision-language
   projector in vision-language MLLMs. To the best of our knowledge, no
   prior research has addressed this critical aspect in the context of
   training foundation models for music understanding. This is
   particularly important because music clips can often be lengthy,
   leading to a large number of music tokens. For instance, the AudioMAE
   encoder outputs 1536 tokens for representing a 30-second music
   clip. While using all available tokens ensures the maximum use of
   music modality information, it may hinder training efficiency and
   limit the utility of the context window of the LLM (2048 in our case).
   In this section, we extensively evaluate the impact of the number of music
   tokens when training \mname.
   
   \figref{img:tokens} displays the training trajectories (log-scale) of
   both Stage (1) and Stage (2) training, where we apply mean-pooling to
   every 2–128 music tokens output by AudioMAE. For instance,
   % alternative version
   mean-pooling every 8 tokens means using only 1536/8 = 192 tokens to represent the 30-second input music clip.
    The difference among mean-pooling 2, 8, and 32 tokens is small, suggesting that
   there may be \emph{redundancies in the representations of the encoded music clip}.
   Although aggressively mean-pooling 128 tokens significantly reduces the overall training time (7.5 hours when pooling 128 tokens vs. 20 hours when mean-pooling 2 tokens), the setting results in a weaker convergence.
   As a result, we empirically focus on model variants with mean-pooling 8 tokens in the next sections to balance between model convergence and training efficiency.
   %----------previous version
   %mean-pooling every 128 tokens means using only 1536/128 = 12 tokens to represent the 30-second input music clip. 
   %It can be observed that aggressively mean-pooling 128 tokens trades performance for training time: 
   %the overall training time is reduced by half (approximately 7.5 hours vs. 20 hours when mean-pooling 2 tokens), but the model's convergence is weaker. 
   %On the other hand, there is no noticeable difference between mean-pooling 2, 8, or 32 tokens, suggesting that
   %\emph{there may be redundancies in the representations of the encoded music clip}.
   %it is reasonable to skip (e.g., by pooling) some unimportant information to speed up the training of MLLMs for music understanding%}. 
   %As a result, we primarily focus on model variants with mean-pooling 8 tokens in the next sections.
   
   \subsection{LoRA, task performance, and music information utility}
   \seclabel{sec:design:lorataskhallu}
   
   \begin{figure}[t]
   \centering
   \includegraphics[width=0.4\textwidth]{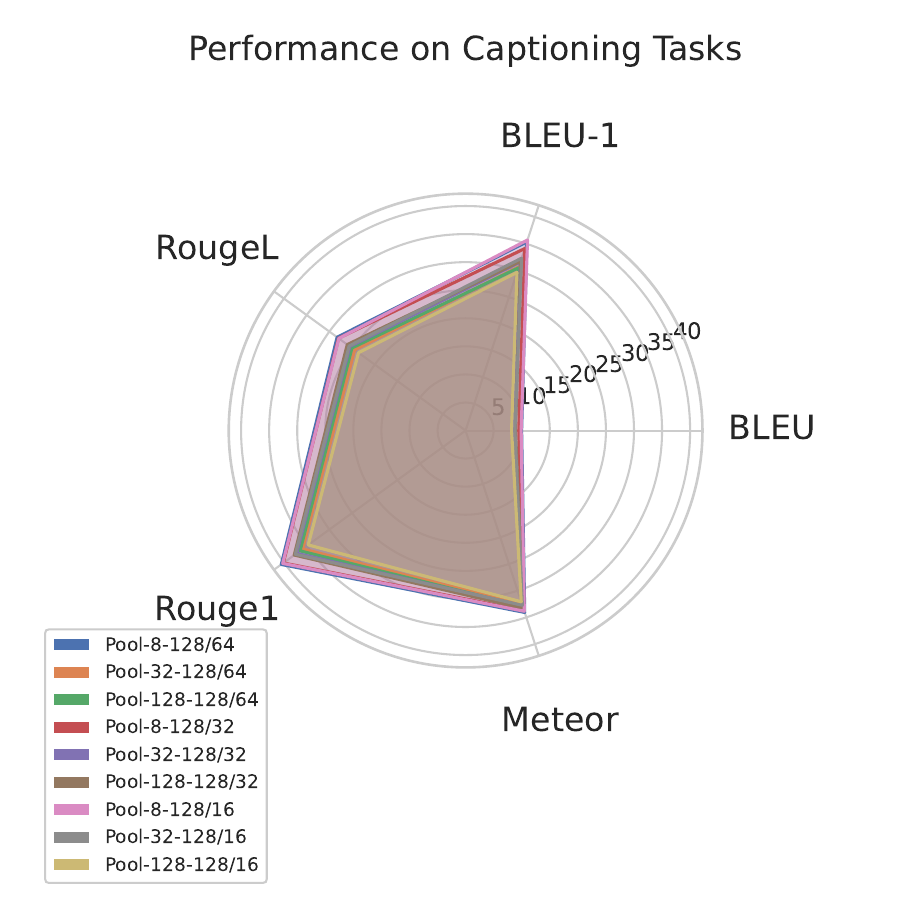}
   \hspace{.5cm}\includegraphics[width=0.4\textwidth]{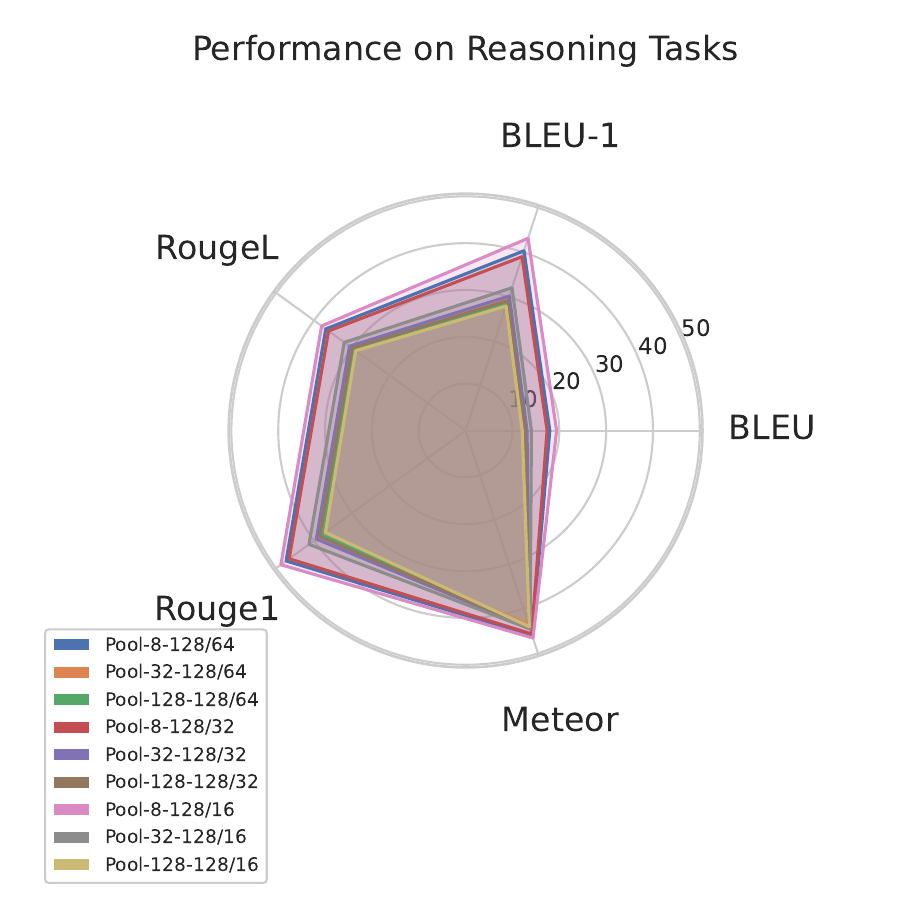}
   \caption{
   Performance of \mname variants on the captioning and reasoning tasks
   of \bname. For each evaluation metric, such as BLEU, we report the
   macro average of the model's performance across all \bname subtasks.
   }
   \figlabel{img:loraablation}
   \vspace{-.5cm}
   \end{figure}

   As introduced in \secref{sec:traindetails}, Low-Rank
   Adaptation (LoRA; \citet{hu2022lora}), is employed in Stage (2)
   training to efficiently adapt \mname's LLM for following instructions
   in the music domain.  Given an LLM weight parameter matrix $\mW \in
   \R^{d\times k}$, instead of directly modifying $\mW$, LoRA introduces
   and trains two matrices, $\mB \in \R^{d\times r}$ and $\mA \in
   \R^{r\times k}$ for adapting $\mW$ to a downstream task:
   $$\mW \leftarrow \mW + \frac{\alpha}{r} \mB\mA .$$
   The LLM weight matrix $\mW$ remains unchanged; the LoRA rank $r$
   determines the number of trainable parameters by controlling the size
   of $\mB$ and $\mA$. The matrix multiplication result, $\mB\mA$,
   represents the changes introduced by adaptation to a downstream task,
   scaled by $\frac{\alpha}{r}$. Here, $\alpha$ is a hyperparameter, and
   typically $r < \alpha$. For \mname, we fix $\alpha = 128$ following
   \citet{liu2023llava,liu2023improvedllava}, while varying the value of
   $r$. Intuitively, a smaller rank $r$ imposes a stricter bottleneck on
   $\mB$ and $\mA$, requiring the learned parameter differences,
   represented by $\mB\mA$, to rely on fewer trainable parameters to
   capture concise and genuine information about the downstream task,
   which are subsequently scaled by a larger $\frac{\alpha}{r}$. In
   contrast, a larger $r$ introduces more trainable parameters, which may
   be prone to learning shortcuts, redundant information, or noise
   \citep{geirhos2020shortcut} during adaptation to the downstream task,
   subsequently scaled by a smaller $\frac{\alpha}{r}$.  In this section,
   we investigate how LoRA configurations affect Stage (2) training, as
   well as reporting the evaluation results of \mname on \bname.

   \textbf{\bname task performance.}
   \figref{img:loraablation} shows the performance of \mname variants on the
   captioning (left) and reasoning (right) tasks of \bname. For each
   evaluation, we report the macro-average performance of each \mname
   variant across all subtasks in \bname. Additionally, \figref{fig:berstcoremucho}
   (left) displays the evaluation results using BertScore
   \citep{bertscore} as the metric. Several observations can be
   made. First, the number of music tokens plays a critical role in task
   performance, echoing the conclusion drawn by \citet{mckinzie2024mm1}.
   Mean-pooling every 8 tokens shows clear advantages over
   32 and 128 tokens, likely due to its preservation of music information.
   However, mean-pooling 32 tokens offers only limited
   improvement over 128 tokens, and the performance decline appears to
   plateau. It is likely that crucial music information is already lost
   when mean-pooling 32 tokens.
   Second, the effectiveness of LoRA parameters show limited impacts on
   task performance, similar to the findings in \citet{ltupaper}.
   \emph{As a result, we will focus on the model variant with mean-pooling 8
   tokens, and LoRA parameters 128/16 in the next sections}.

   \textbf{Music information utility.}  
   Given the large-scale pretraining data of the LLM, which already contains rich knowledge about music, an MLLM may be able to answer questions about music without relying on the information in the music clip. 
   Hence, this section addresses a key question: Does \mname genuinely utilize information from the input clip to understand the music more effectively? 
   To investigate this, 
   we evaluate \mname variants on MuChoMusic
   \citep{weck2024muchomusic}, a dataset containing multiple-choice
   questions focused on music understanding. 
   %For example, 
   Questions such
   as ``Which sub-genre of rock music would best classify this piece?''
   require the model to select the correct option from four
   candidates. Notably, such questions could %potentially 
   be answered
   based on the most common or probable sub-genre from the LLM’s
   pretraining data, allowing the model to perform reasonably well
   without actually relying on the music input.
   \figref{fig:berstcoremucho} (right) presents the MuChoMusic results of
   \mname variants. The ``No Music'' condition refers to replacing the
   input music clip with a white noise clip, while ``OpenMU'' displays the
   results when actual music clips are used. It is evident that music
   information is crucial for \mname to achieve strong performance;
   \mname effectively utilizes music information
   rather than relying on shortcuts \citep{geirhos2020shortcut}.

   \begin{figure}[t] \centering
     \includegraphics[width=.75\textwidth]{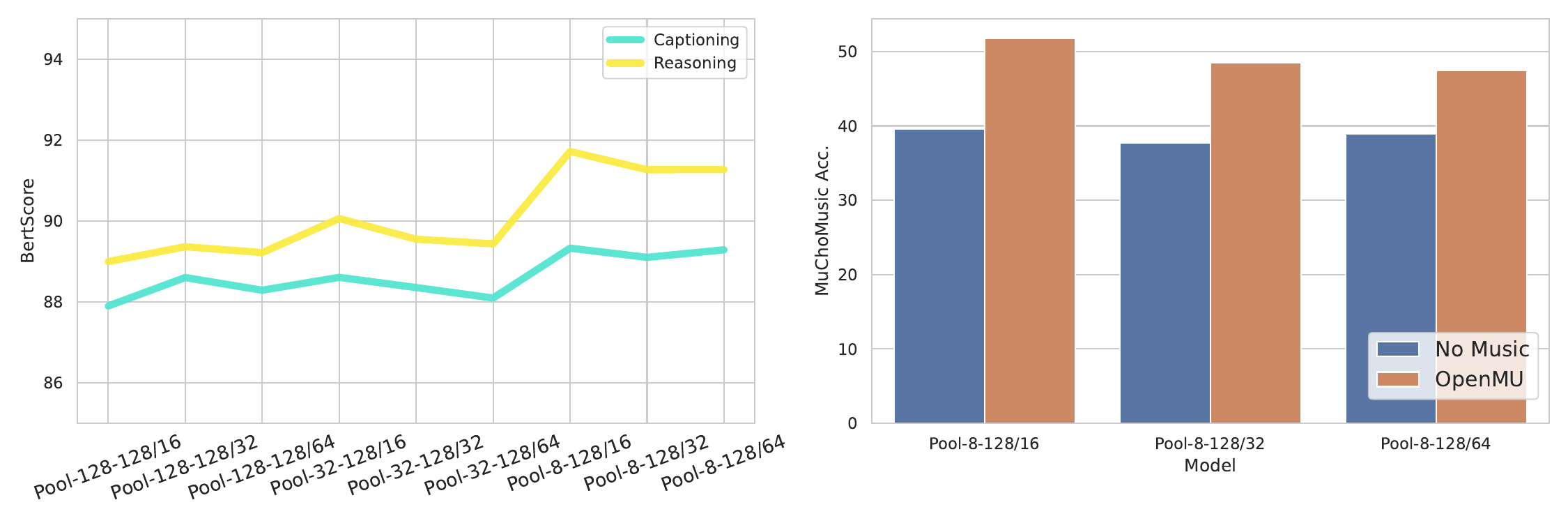}
     \caption{Left: Performance of \mname variants on the captioning and
     reasoning tasks of \bname using BertScore as the metric. Right:
      \mname performance on MuChoMusic.}
     \figlabel{fig:berstcoremucho}
     \vspace{-.3cm}
   \end{figure}
   
   \section{Overall results}
   \vspace{-.3cm}
   \seclabel{sec:finalres}
   \begin{table}[t]
\centering
\scriptsize
\rastrech{1.3}
\begin{tabular}{r|cc|cc|cc|cc|cc|cc|}
\cline{2-13}
\multicolumn{1}{l|}{}                       & \multicolumn{2}{c|}{\textbf{BLEU-1}} & \multicolumn{2}{c|}{\textbf{BLEU}} & \multicolumn{2}{c|}{\textbf{Rouge1}} & \multicolumn{2}{c|}{\textbf{RougeL}} & \multicolumn{2}{c|}{\textbf{BertScore}} & \multicolumn{2}{c|}{\textbf{Meteor}} \\ \cline{2-13}
\multicolumn{1}{c|}{}                       & OMU            & MUL             & OMU           & MUL            & OMU            & MUL             & OMU            & MUL             & OMU             & MUL               & OMU            & MUL             \\ \hline
\multicolumn{1}{|r|}{\textbf{MusicCaps}}    & \textbf{25.62}    & 9.78             & \textbf{2.89}    & 0.73            & \textbf{27.99}    & 21.22            & \textbf{18.56}    & 15.64            & 86.63              & \textbf{86.85}     & \textbf{21.83}    & 11.03            \\ \hline
\multicolumn{1}{|r|}{\textbf{MI-short}}     & 18.92             & \textbf{50.01}   & 8.75             & \textbf{23.90}  & 41.03             & \textbf{52.69}   & 37.75             & \textbf{47.73}   & 89.90              & \textbf{92.83}     & 43.81             & \textbf{49.10}   \\ \hline
\multicolumn{1}{|r|}{\textbf{MI-long}}      & \textbf{36.66}    & 2.13             & \textbf{4.18}    & 0.18            & \textbf{38.74}    & 19.34            & \textbf{21.70}    & 13.55            & \textbf{87.24}     & 85.90              & \textbf{24.43}    & 8.83             \\ \hline
\multicolumn{1}{|r|}{\textbf{LPMusicMTT}}   & \textbf{23.83}    & 18.87            & \textbf{2.56}    & 0.56            & \textbf{29.23}    & 21.68            & \textbf{21.57}    & 16.07            & \textbf{89.75}     & 88.37              & \textbf{24.60}    & 13.96            \\ \hline
\multicolumn{1}{|r|}{\textbf{Music4all}$^*$}    & \textbf{51.03}    & 5.11             & \textbf{18.69}   & 0.36            & \textbf{51.31}    & 19.51            & \textbf{34.80}    & 14.31            & \textbf{91.04}     & 86.64              & \textbf{43.58}    & 10.07            \\ \hline
\multicolumn{1}{|r|}{\textbf{MusicQA-Test}} & 19.60             & \textbf{19.65}   & 2.22             & \textbf{5.08}   & 23.96             & \textbf{31.74}   & 17.39             & \textbf{28.07}   & 87.30              & \textbf{89.45}     & \textbf{26.34}    & 21.73            \\ \hline
\multicolumn{1}{|r|}{\textbf{GTZAN}$^*$}        & \textbf{45.38}    & 4.56             & \textbf{11.78}   & 0.32            & \textbf{44.01}    & 20.41            & \textbf{27.62}    & 15.17            & \textbf{89.34}     & 87.03              & \textbf{34.71}    & 10.65            \\ \hline
\multicolumn{1}{|r|}{\textbf{MusicNet}$^*$}     & \textbf{52.68}    & 1.17             & \textbf{22.14}   & 0.00            & \textbf{56.44}    & 14.49            & \textbf{38.11}    & 12.25            & \textbf{91.97}     & 85.59              & \textbf{45.92}    & 6.56             \\ \hline
\multicolumn{1}{|r|}{\textbf{MTG-Jamendo}$^*$}  & \textbf{47.56}    & 5.32             & \textbf{15.83}   & 0.41            & \textbf{49.66}    & 20.29            & \textbf{33.76}    & 15.94            & \textbf{90.79}     & 87.72              & \textbf{39.44}    & 10.06            \\ \hline
\end{tabular}
\caption{\bname captioning results (in \%) of \mname (OMU) and \mul (MUL).}
\tablabel{vsmullamacaption}
\vspace{-.5cm}
\end{table}

   \begin{table}[bt]
\centering
\scriptsize
\rastrech{1.3}
\begin{tabular}{r|cc|cc|cc|cc|cc|cc|}
\cline{2-13}
\multicolumn{1}{l|}{}                       & \multicolumn{2}{c|}{\textbf{BLEU-1}} & \multicolumn{2}{c|}{\textbf{BLEU}} & \multicolumn{2}{c|}{\textbf{Rouge1}} & \multicolumn{2}{c|}{\textbf{RougeL}} & \multicolumn{2}{c|}{\textbf{BertScore}} & \multicolumn{2}{c|}{\textbf{Meteor}} \\ \cline{2-13}
\multicolumn{1}{c|}{}                       & OMU               & MUL              & OMU              & MUL             & OMU               & MUL              & OMU               & MUL              & OMU                & MUL                & OMU               & MUL              \\ \hline
\multicolumn{1}{|r|}{\textbf{Music4all}$^*$}    & \textbf{49.20}    & 18.13            & \textbf{23.31}   & 5.97            & \textbf{53.26}    & 34.11            & \textbf{41.08}    & 25.01            & \textbf{92.34}     & 89.63              & \textbf{49.96}    & 22.51            \\ \hline
\multicolumn{1}{|r|}{\textbf{MusicQA-Test}} & 24.84             & \textbf{40.64}   & 9.46             & \textbf{22.47}  & 35.86             & \textbf{51.29}   & 30.66             & \textbf{47.54}   & 89.70              & \textbf{92.59}     & 40.04             & \textbf{46.15}   \\ \hline
\multicolumn{1}{|r|}{\textbf{GTZAN}$^*$}        & \textbf{50.26}    & 16.16            & \textbf{22.07}   & 5.95            & \textbf{52.96}    & 35.18            & \textbf{38.57}    & 26.20            & \textbf{92.02}     & 89.89              & \textbf{46.87}    & 21.84            \\ \hline
\multicolumn{1}{|r|}{\textbf{MTT}$^*$}          & \textbf{45.52}    & 21.70            & \textbf{21.18}   & 8.31            & \textbf{50.83}    & 38.70            & \textbf{39.93}    & 29.92            & \textbf{92.03}     & 90.63              & \textbf{48.06}    & 26.33            \\ \hline
\multicolumn{1}{|r|}{\textbf{MTG-Jamendo}$^*$}  & \textbf{45.87}    & 23.69            & \textbf{21.12}   & 8.59            & \textbf{50.78}    & 38.45            & \textbf{39.74}    & 29.27            & \textbf{92.01}     & 90.47              & \textbf{47.97}    & 26.42            \\ \hline
\end{tabular}
\caption{\bname reasoning results (in \%) of \mname (OMU) and \mul (MUL).}
\tablabel{vsmullamareason}
\end{table}

   In this section, we compare \mname with \mul, a widely used music
   understanding model, on \bname. For \mname, we use the variant of
   mean-pooling 8 tokens with LoRA parameters 128/16. For \mul, we use
   the checkpoint released by \citet{mullama}.

   \textbf{Music captioning and reasoning} results are presented in
   \tabref{vsmullamacaption} and \tabref{vsmullamareason},
   respectively. We observe that \mname consistently outperforms \mul
   across various captioning and reasoning tasks. Interestingly, \mul
   lags behind \mname on MusicCaps, despite the fact that the MusicCaps
   test set was used during \mul's pretraining stage \citep{mullama,
   deng-etal-2024-musilingo}.  We believe this is due to the small size
   of MusicCaps—its effectiveness was likely overshadowed by the larger
   finetuning datasets used for \mul.
   
   \mul outperforms \mname on the MusicInstruct-short captioning task \citep{deng-etal-2024-musilingo} and the MusicQA-test reasoning task \citep{mullama} in terms of surface form matching metrics by a large margin. 
   However, we found that the gold references in these two subsets are biased to contain a large portion of repeated parts from the questions. 
   For example, in MusicQA-test reasoning, to the question ``What is \underline{the alternative genre of music in the audio}?'', the gold standard reference is ``\underline{The alternative genre of music in the audio} is postrock.'' 
   This observation is further supported by objective metrics. Compared to the edit distance of 225 and the Jaccard similarity score of 23.9\% for MTT, 
   MusicQA-test reasoning has an edit distance of 90 and a Jaccard similarity score of 36.2\%.  
   Since \mul tends to repeat the question before providing an answer, the behavior might have inflated the surface-level form matching scores in these subsets. 
   As a result, we recommend practitioners downweight these subsets when evaluating music understanding models.

   \begin{table}[t]
   \centering
   \scriptsize
   \rastrech{1.3}
   \begin{tabular}{r|ccc|c|}
   \cline{2-5}
   \multicolumn{1}{c|}{}           & \multicolumn{3}{c|}{\textbf{Accuracy}} & \textbf{IFR}   \\ \cline{2-5}
   \multicolumn{1}{c|}{}           & All   & Knowledge & Reasoning & All   \\ \hline
   \multicolumn{1}{|r|}{\textbf{MusiLingo}} & 21.1  & 22.0      & 19.2      & 71.6  \\
   \multicolumn{1}{|r|}{\textbf{\mul}}   & 32.4  & 32.3      & 31.3      & 79.4  \\
   \multicolumn{1}{|r|}{\textbf{M2UGen}}    & 42.9  & 44.9      & 41.2      & \textbf{96.4}  \\
   \multicolumn{1}{|r|}{\textbf{OpenMU}}    & \textbf{51.8}  & \textbf{51.4}      & \textbf{51.4}      & 94.8  \\ \hdashline
   \multicolumn{1}{|r|}{Random}    & 25.0  & 25.0      & 25.0      & 100.0 \\ \hline
   \end{tabular}
   \caption{MuChoMusic accuracy and instruction-following rate
   (IFR) of \mname and prior music understanding models. Numbers are in \%. 
   MuChoMusic contains multiple-choice
   questions; MusiLingo, \mul, M2UGen performances are from \citet{weck2024muchomusic}.
   ``Random'' shows random guessing results. We assume ``Random'' will always select an option, hence its IFR is 100\%.
   }
   \tablabel{tab:mucho}
   \vspace{-5mm}
   \end{table}

   \begin{table}[h!]
   \centering
   \scriptsize
   \rastrech{1.3}
   \hspace{-1.5cm}
   \begin{minipage}{.6\linewidth}
   \centering
   \begin{tabular}{c|c|c|c|c|c|c|}
   \cline{2-7}
                                     & BLEU-1         & BLEU          & Rouge1         & RougeL         & BertScore      & Meteor         \\ \hline
   \multicolumn{1}{|c|}{BART-Fusion} & \textbf{25.79} & \textbf{6.48} & \textbf{32.18} & \textbf{17.99} & 83.03          & \textbf{27.97} \\ \hline
   \multicolumn{1}{|c|}{OpenMU}      & 25.60          & 5.19          & 31.31          & 17.03          & \textbf{83.14} & 27.01          \\ \hline
   \end{tabular}
   \end{minipage}
   \hspace{0.1\linewidth}
   \begin{minipage}{.1\linewidth}
   \centering
   \vspace{-.4cm}
   \begin{tabular}{|c|c|}
   \hline
   Chords    & 94.95 \\ \hline
   Tempo     & 95.83 \\ \hline
   Key       & 100   \\ \hline
   Downbeats & 100   \\ \hline
   \end{tabular}
   \end{minipage}
   \caption{Lyrics understanding results (left) and tool using accuracy (right). Numbers are in \%.}
   \tablabel{tab:lyricstool}
   \vspace{-.3cm}
   \end{table}

   \textbf{Multiple-choice questions}. We compare \mname with \mul, along with other
   available music understanding models, on the multiple-choice question
   dataset MuChoMusic. \citet{deng-etal-2024-musilingo} is a concurrent
   work to \mul while M2UGen \citep{hussain2023m} adds music generation
   ability to \mul. \tabref{tab:mucho} shows that \mname achieves
   state-of-the-art music understanding performance on MuChoMusic.
   
   \textbf{Lyrics understanding.} \tabref{tab:lyricstool} (left) compares
   \mname's performance with BART-fusion \citep{lyricsbart}, a model
   specifically designed for lyrics understanding. For simplicity, we
   reuse the same hyperparameters from Stage (2) training, except for
   extending the training to 20 epochs. \mname outperforms BART-fusion in
   BertScore but slightly lags behind on other metrics. Future models
   could explore further hyperparameter tuning or architectural
   modifications to improve performance.  \textbf{Tool using accuracy}.
   \tabref{tab:lyricstool} (right) reports the accuracy of \mname when
   calling external MIR tools. We consider an exact match as a hit. For
   example, in chords estimation, if the gold reference is
   ``[GetMusicChords(10, 20)]'', the model must accurately output the type and arguments of the tool
   call to be considered a hit. Extra calls are considered a miss. As
   expected, \mname performs well on this task and learns to call MIR
   tools effectively. It is promising to integrate more MIR tools to
   handle a broader range of task types and complexities.
   
   \section{Conclusion}
   We presented \bname, a large-scale benchmark suite containing
   approximately one million examples for training and evaluating LLM-based 
   music understanding models. We construct \bname by creating new annotations 
   as well as leveraging existing datasets.
   We trained our music understanding model, \mname, with
   extensive ablations and demonstrated that it outperforms baseline
   models such as \mul. Both \mname and \bname are open-sourced to
   facilitate future research in music understanding and enhance the
   efficiency of creative music production. Future work may explore
   extending \mname to support multiple music clips as input and enable
   in-context learning for music understanding. Another promising
   direction is enabling \mname to integrate more MIR tools, combining
   the strengths of LLMs and established tools for deeper music
   understanding.

\bibliography{iclr2025_conference}
\bibliographystyle{iclr2025_conference}

%\clearpage
\appendix
%\clearpage
\section{Appendix}
\subsection{Training details and hyperparameters}
\seclabel{apdx:hypers}

In this section, we describe the detailed settings and hyperparameters
used for training \mname.

All experiments were conducted using 8-16 A100 40GB GPUs, with BF16
enabled to ensure stable training. We use DeepSpeed ZERO-3
\citep{rajbhandari2020zero} and Flash Attention 2
\citep{dao2022flashattention} to reduce the memory consumption.  We
utilized the Adam optimizer and a cosine learning rate scheduler, with
a 30\% warm-up ratio.

For Stage (1) training, we pretrained \mname for 15 epochs on the
captioning subtask of \bname, which consists of approximately 275K
pairs of music clips and corresponding captions. Stage (1) training
took approximately 10 hours for the checkpoint we evaluated (i.e.,
mean-pooling every 8 music tokens as illustrated in
\figref{img:tokens}. The initial learning rate was set to 1e-3, with a
batch size of 8 per GPU.

For Stage (2) training, we extended pretraining of \mname for 10 epochs
on the captioning and reasoning subtasks of \bname, which comprise
roughly one million training examples. The initial learning rate was
set to 2e-5, with the same per-GPU batch size of 8. Stage (2) required
approximately 40 hours due to the increased size of training data.

For the lyrics understanding subtask, we trained \mname for 20 epochs,
reusing the hyperparameters from Stage (2). Similarly, for the tool
using subtask, we reused the Stage (2) hyperparameters but reduced the
number of epochs to 5 due to the smaller dataset size for this task.

\subsection{Metadata of datasets}
\seclabel{apdx:metadata}
In this paper, we contribute to creating the large-scale benchmark
suite \bname for music understanding.

In contrast to other modalities such as images, where rich natural
language descriptions are widely available across the internet
\citep{schuhmann2022laionb}, music clips are often accompanied by
tags, such as genre, year, and instruments. We consider these tags to
be a form of metadata for the music clips. When constructing \bname,
we bootstrap captions and reasoning texts in natural language about
the music clips based on this metadata by prompting GPT-3.5.

\begin{table}[t]
\centering
\scriptsize
\rastrech{1.3}
\begin{tabular}{c|c|c|c|c|c|c|c|c|}
\cline{2-9}
                                           & \textbf{Tempo} & \textbf{Energy} & \textbf{Valence} & \textbf{Danceability} & \textbf{Genre} & \textbf{Mood} & \textbf{Instrument}  & \textbf{Others} \\ \hline
\multicolumn{1}{|c|}{\textbf{Music4all}}   & $\bigcirc$     & $\bigcirc$      & $\bigcirc$       & $\bigcirc$            & $\triangle$    & $\triangle$   & $\triangle$ & $\triangle$        \\ \hline
\multicolumn{1}{|c|}{\textbf{GTZAN}}       & $\bigcirc$     & $\times$        & $\times$         & $\times$              & $\bigcirc$     & $\times$      & $\times$    & $\times$           \\ \hline
\multicolumn{1}{|c|}{\textbf{MusicNet}}    & $\times$       & $\times$        & $\times$         & $\times$              & $\times$       & $\times$      & $\bigcirc$  & $\times$            \\ \hline
\multicolumn{1}{|c|}{\textbf{MTT}}         & $\times$       & $\times$        & $\times$         & $\times$              & $\triangle$    & $\triangle$   & $\triangle$ & $\triangle$        \\ \hline
\multicolumn{1}{|c|}{\textbf{MTG-Jamendo}} & $\times$       & $\times$        & $\times$         & $\times$              & $\triangle$    & $\triangle$   & $\triangle$ & $\triangle$        \\ \hline
\end{tabular}
\caption{The metadata associated with each subtask dataset in
\bname.
$\bigcirc$: the metadata, e.g., tempo, is available for the music clips.
$\triangle$: the metadata \emph{maybe} available for some music clips, but not for all of them.
$\times$: the metadata is not available for the music clips.
}
\tablabel{tab:metadata}
\end{table}

% \begin{table}[]
% \centering
% \scriptsize
% \rastrech{1.3}
% \begin{tabular}{c|c|c|c|c|c|c|c|}
% \cline{2-8}
%                                            & \textbf{Tempo} & \textbf{Energy} & \textbf{Valence} & \textbf{Danceability} & \textbf{Genre} & \textbf{Mood} & \textbf{Instrument} \\ \hline
% \multicolumn{1}{|c|}{\textbf{Music4all}}   & O              & O               & O                & O                     & Δ              & Δ             & Δ                   \\ \hline
% \multicolumn{1}{|c|}{\textbf{GTZAN}}       & O              & X               & X                & X                     & O              & X             & X                   \\ \hline
% \multicolumn{1}{|c|}{\textbf{MusicNet}}    & X              & X               & X                & X                     & X              & X             & O                   \\ \hline
% \multicolumn{1}{|c|}{\textbf{MTT}}         & X              & X               & X                & X                     & Δ              & Δ             & Δ                   \\ \hline
% \multicolumn{1}{|c|}{\textbf{MTG-Jamando}} & X              & X               & X                & X                     & Δ              & Δ             & Δ                   \\ \hline
% \end{tabular}
% \tablabel{tab:metadata}
% \end{table}

\tabref{tab:metadata} demonstrates the various types of metadata used
in the \bname subtasks to create music understanding examples. Due to
the broad coverage, music clips from different \bname subsets are
associated with diverse types of metadata. Even within the same
subtask, different music clips may be tagged with only a limited set
of metadata types. We detail how we process the metadata of each music
clip as follows.

\begin{figure}[t]
\centering
\includegraphics[width=\textwidth]{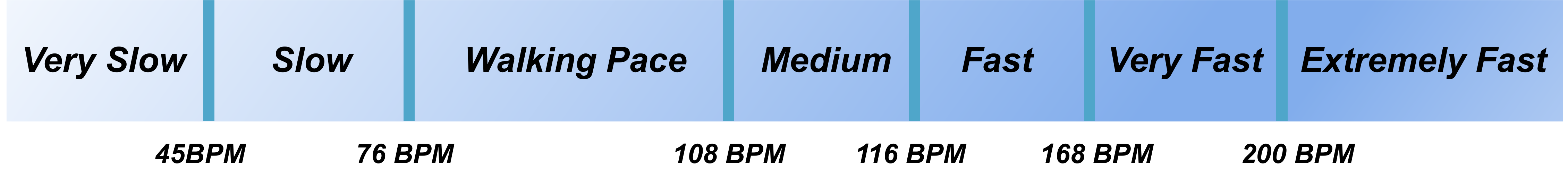}
\caption{
Converting numerical values (in beats per minute; BPM) of music tempo
to natural language descriptions.  The conversion is done based on the
Italian musical terms.
}
\figlabel{fig:bpmrange}
\end{figure}

\textbf{Tempo}. Music clips in two datasets, music4all and GTZAN, are
associated with tempo, and we convert the numerical values into
natural language descriptions, following the Italian musical
terms\footnote{Italian musical terms:
\url{https://www.musicca.com/musical-terms}.} as shown in image
\figref{fig:bpmrange}.

For \textbf{energy}, \textbf{valence}, \textbf{danceability}, which
are float scores ranging from 0 to 1, we convert them into natural
language descriptions using empirical thresholds of 0.3 and 0.7. Taking
energy as an example, we consider a energy level $s$, where $s\geq$0.7
as a high energy level, 0.7$>s\geq$0.3 as a medium energy level, and
0.3$>$s as a low energy level.

For \textbf{genre}, \textbf{mood}, and \textbf{instrument} of
Music4all, MTT, MTG-Jamendo, we merge their original metadata by
manual annotations and corrections for consistency. Concretely, we
keep the top 50 tags of MTT and MTG-Jamendo, following the
recommendation of the authors \citep{mttdataset,mtgdataset}, and use
the top 166 tags of Music4all, as recommended by
\texttt{music4all\_contrib}\footnote{\url{https://github.com/keunwoochoi/music4all_contrib}}.
The corrections involves actions such as de-compounding
(``acousticguitar'' $\rightarrow$ ``acoustic guitar''), unifying
(``Female vocalists'' $\rightarrow$ ``female vocal''), expanding
(``synth'' $\rightarrow$ ``synthesizer'') the tags, and the resulting
metadata tags are list as follows:

\begin{tcolorbox}[colback=white, colframe=black, width=\textwidth,
arc=0mm, boxrule=0.3mm,title=\hspace{3cm}{Metadata of genre, instrument, mood, and others}]
\begin{center}
  \scriptsize
  \verbatiminput{imgs/merged_metadata.txt}
\end{center}
\end{tcolorbox}

As a result, a JSON formatted metadata is created for each of the music clips:
\begin{verbatim}
{
    "dataset_name": "music4all",
    "audio_filename": "4MqXFtyr1XwxrShX.mp3",
    "tempo": "walking pace tempo",
    "valence": "medium valence",
    "energy": "high energy",
    "danceability": "medium danceable",
    "genre": [
      "rock",
      "pop",
      "electronic"
    ],
    "mood": [
      "ambient"
    ]
}
\end{verbatim}

which is then employed to prompt GPT3.5 to create examples for music
understanding, as described in the next section \secref{apdx:prompts}.

\subsection{Prompts and dataset format}
\seclabel{apdx:prompts}

Based on the metadata of each music clip (\secref{apdx:metadata}), we prompt
GPT-3.5 to generate examples for the music understanding tasks. Our
prompts are adapted from those used by \citet{gardner2023llark}, with
modifications tailored to the available metadata of different \bname
subsets and subtasks. As an example, we present the prompt used for
the music captioning task in Music4All:

\begin{tcolorbox}[colback=white, colframe=black, width=\textwidth, arc=0mm, boxrule=0.3mm, title=\hspace{4cm}Example prompt used for Music4all]
\begin{center}
  \scriptsize
  \verbatiminput{imgs/music4all_captioning_prompt.txt}
\end{center}
\end{tcolorbox}

In short, we ask GPT-3.5 to generate an informative caption that can
describe the key features and components of the music clip, based on
its internal music knowledge accumulated during pretraining and our
provided rich metadata. The generated results of the
example in \secref{apdx:metadata} is:

\begin{small}
\begin{verbatim}
{
    "instruction": "Explain the contents of this song.",
    "output": "This music clip features high energy with a
        walking pace tempo. It combines elements from rock,
        pop, and electronic genres, creating a dynamic and lively
        soundscape. The medium valence contributes to a balanced
        emotional tone, while the medium danceability suggests a
        rhythmic and engaging quality, suitable for movement and dance.
        The ambient mood adds a layer of atmospheric texture to the
        overall vibrant and energetic composition.",
    "local_audio_path": "4MqXFtyr1XwxrShX.mp3",
    "task": "captioning",
    "dataset": "music4all_test"
},
\end{verbatim}
\end{small}

which is then leveraged to train or test \mname according to the dataset split.

\subsection{Dataset splits}
\seclabel{apdx:splits}
In this section, we provide details about the train/test splits of the
\bname subtasks. Specifically, for MusicCaps, MusicInstruct,
LPMusicCaps, LPMusicMTT, MusicQA, MusicNet, BART-Fusion,
and MuchoMusic, we follow the train/test splits proposed in the
original papers. 
For GTZAN, we used the widely accepted filter-fault split~\citep{kereliuk2015gtzan_split}, 
and the split from MARBLE~\citep{yuan2023marble} for MTT.

For Music4All, we start with the 800 music clips from BART-Fusion as
the initial test set. We then expand this set by randomly sampling
music clips until the total reaches 5,000. The remaining music clips
and their annotations are used as training data.
For MTG-Jamendo, we use annotations where the music clips from folds
90 to 99 of the original dataset \citep{mtgdataset} serve as the test
data, while the remaining clips and their annotations are treated as
training data.
For tool using, we randomly sample 80\% examples for training and 20\%
for testing.

\subsection{Tools}
\seclabel{apdx:tools}
We define simple tools for solving MIR tasks such as tempo estimator.
They are implemented as simple Python wrapper to the 
\texttt{Madmom} toolkit \citep{madmom}, which has been widely used 
in MIR. For example, the tempo estimator can be 
implemented\footnote{
In our implementation, we use pseudo names, such as \texttt{F1}, for the tools. 
We found that Llama3 tends to hallucinate new tools when camel case names like "EstimateTempo" are used, 
likely due to the presence of code in its pretraining data \citep{llama3}.
} as:

\begin{small}
\begin{verbatim}
from madmom.features.beats import RNNBeatProcessor
from madmom.features.tempo import TempoEstimationProcessor


def EstimateTempo():
    wav = load_audio(AUDIO_FILE)
    beat_proc = RNNBeatProcessor()
    tempo_proc = TempoEstimationProcessor(fps=100)
    beat_acts, tempo_acts = beat_proc(wav), tempo_proc(beat_acts)
    tempo_est = round(tempo_acts[0][0], 1)
    return tempo_est
\end{verbatim}
\end{small}

\mname then calls for such a tool to estimate the tempo, when being asked questions
such as ``Let me know the tempo of this music clip.'' and replying with 
``The music has tempo [EstimateTempo() $\rightarrow$ $n$] beats per minute.''.
\end{document}